\documentclass{PoS}

\title{A QCD analysis of ZEUS diffractive data}

\ShortTitle{A QCD analysis of ZEUS diffractive data}

\author{\speaker{Matthew Wing} \thanks{I am grateful to Aharon Levy who stepped in to give the talk as I could not attend.  My sympathies lie with the organisers of the conference who must have endured a testing time due to the disturbance caused by a volcano eruption; I am sorry that I could not come because of this.} \ (On behalf of the ZEUS Collaboration)%

\\

        Department of Physics and Astronomy, UCL, Gower Street, London WC1E 6BT, UK\\ 

        E-mail: \email{mw@hep.ucl.ac.uk}}

\abstract{ZEUS inclusive diffractive cross-section measurements have been used in a 
          next-to-leading-order QCD analysis to extract the diffractive parton 
          distribution functions.  Data on diffractive dijet production in deep 
          inelastic scattering have also been used to constrain the gluon density.  
          Predictions based on the extracted parton densities are able to describe 
          measurements of dijet photoproduction.}

\FullConference{XVIII International Workshop on Deep-Inelastic Scattering and Related Subjects\\
                 April 19 -23, 2010\\
                 Convitto della Calza, Firenze, Italy}

\begin{document}

\newcommand{\xpom}{x_{_{I\!\!P}}}
\newcommand{\fpom}{f_{_{I\!\!P}}}
\newcommand{\freg}{f_{_{I\!\!R}}}
\newcommand{\fregi}{f_i^{{I\!\!R}}}

\section{Introduction}

Diffractive deep inelastic scattering (DIS) presents a very striking signature in which the 
proton does not break up, a large (rapidity) gap of inactivity is present in the direction 
of the proton and a high energy scattered electron is observed.  Such processes, in which 
the four-momentum exchange, $Q^2$, of the virtual photon is sufficiently high, are 
amenable to predictions of perturbative QCD.  The extent to which diffraction can be 
described by QCD is investigated here; as diffraction represents about 10\% of the total 
DIS cross section, it is essential to understand its nature.  Diffraction, according to 
the factorisation theorem~\cite{fact}, can be described as a convolution of a hard scattering 
process and diffractive parton density functions (DPDFs) which describe the densities 
of partons in the proton in a process containing a fast proton in the final state.  
Additionally, Higgs Bosons may be produced (at the LHC) via a diffractive process in which 
fast protons are detected; a deeper understanding of diffraction could therefore aid in  
the discovery of the Higgs Boson.

The extraction of DPDFs has been performed by several groups~\cite{H1DPDF1,H1DPDF2,MRW-GBL} 
which have been able to successfully describe diffractive DIS data.  Jet data have also been 
used~\cite{H1DPDF2} in fits to improve the gluon density.  However, the DPDFs can not 
describe Tevatron data~\cite{cdf} on diffraction, being about an order of magnitude higher 
that the data.  This apparent factorisation breaking can be roughly described by invoking 
models of secondary scatters which destroy the rapidity 
gap of inactivity~\cite{kkmr}.  Therefore the thrust of these proceedings~\cite{fit-paper} 
is to use the most precise ZEUS data on diffractive DIS and jet production in diffractive 
DIS~\cite{fit-incl,fit-jet} and improved theoretical assumptions to provide the best DPDFs possible 
to both test pQCD and to be used for predicting other processes.

\section{Fitting framework and procedure}

The DPDFs are parametrised as a function of the hard scale, $Q^2$, and the 
longitudinal momentum fraction, $z$, of the parton entering the hard sub-process.  This assumes 
proton-vertex factorisation, used to model the dependence on the fraction of 
the momentum of the proton carried by the diffractive exchange, $\xpom$.  Two contributions 
were assumed, called Pomeron and Reggeon, separately factorisable into a term depending only 
on $\xpom$ and a term depending only on $z$ and $Q^2$, 
\[
f_i^D(z, \xpom; Q^2) = \fpom (\xpom) f_i(z,Q^2) + \freg (\xpom) \fregi(z,Q^2)\,,
\]
where the Reggeon parton densities, $\fregi(z,Q^2)$, were taken from a parametrisation 
derived from fits to pion structure-function data~\cite{pionSF} and the Pomeron, 
$\fpom (\xpom)$~\cite{fit-incl}, 
and Reggeon, $\freg (\xpom)$~\cite{pomreg}, fluxes were taken from elsewhere.
Next-to-leading-order (NLO) DGLAP QCD theory was fit to the data by minimising a 
$\chi^2$ using the ``offset'' method.  The renormalisation and factorisation scales were 
both set to $Q$ when fitting the inclusive data and to $Q$ and the transverse jet energy, 
$E_T^{\rm jet}$, when fitting the jet data.  The general-mass variable-flavour number 
scheme~\cite{tr} was used to account for charm and bottom of masses 1.35 
and 4.3 GeV.  The starting scale was $Q_0^2 = 1.8$\,GeV$^2$ with strong coupling constant,  
$\alpha_s(M_Z) = 0.118$.

Three different fits were performed : two to the inclusive data, fits ``S'' and ``C'', but 
with different parametrisations for the gluon density, $A_g\,z^{B_g}\,(1-z)^{C_g}$ or $A_g$, 
respectively; and one fitting both the inclusive and jet data, fit ``SJ'', with the same 
parametrisation as fit S.

\section{Fit results}

\begin{figure}[htp]
\begin{center}
\includegraphics[width=0.6\textwidth]{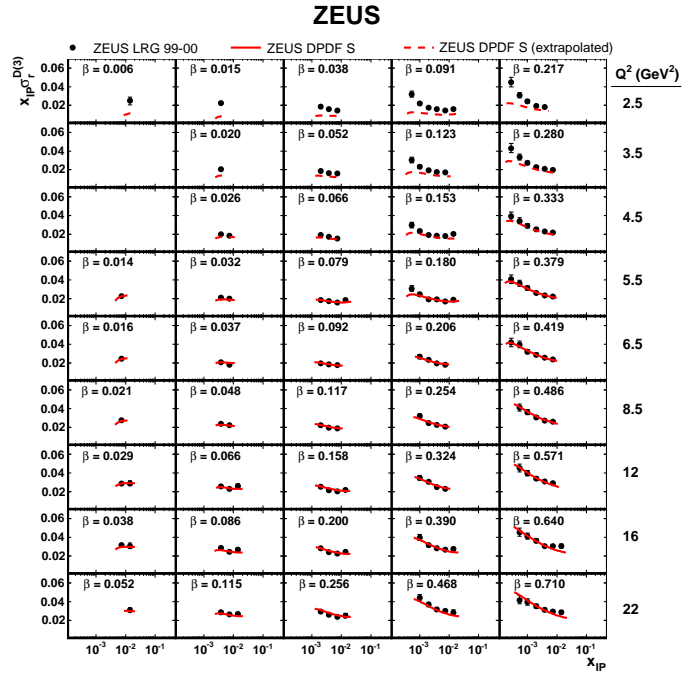}
\vspace{-0.3cm}
\caption{Comparison of NLO QCD DPDF fits to inclusive diffractive DIS data at low $Q^2$.}
\label{fig:incl}
\end{center}
\end{figure}

Results of the NLO QCD fit to the high-statistics inclusive diffractive DIS data at low 
$Q^2$ are shown in Fig.~\ref{fig:incl} as a function of $\xpom$ for different values of the 
Bjorken variable, $\beta$.  Only data above 5\,GeV$^2$ were used in the fit and 
are described well; data below 5\,GeV$^2$ gave a poor fit and could not be described 
within this framework.  Data at higher $Q^2$ are similarly well described.  Complimentary 
data, where the proton was tagged event-by-event rather than statistically, 
determined by a lack of activity in the direction of the proton, were also well 
described, although they are of lower precision and hence do not constrain the fit 
with the same power as those of Fig.~\ref{fig:incl}.

\vspace{-0.3cm}

\begin{figure}[htp]
\begin{center}
\includegraphics[width=0.49\textwidth]{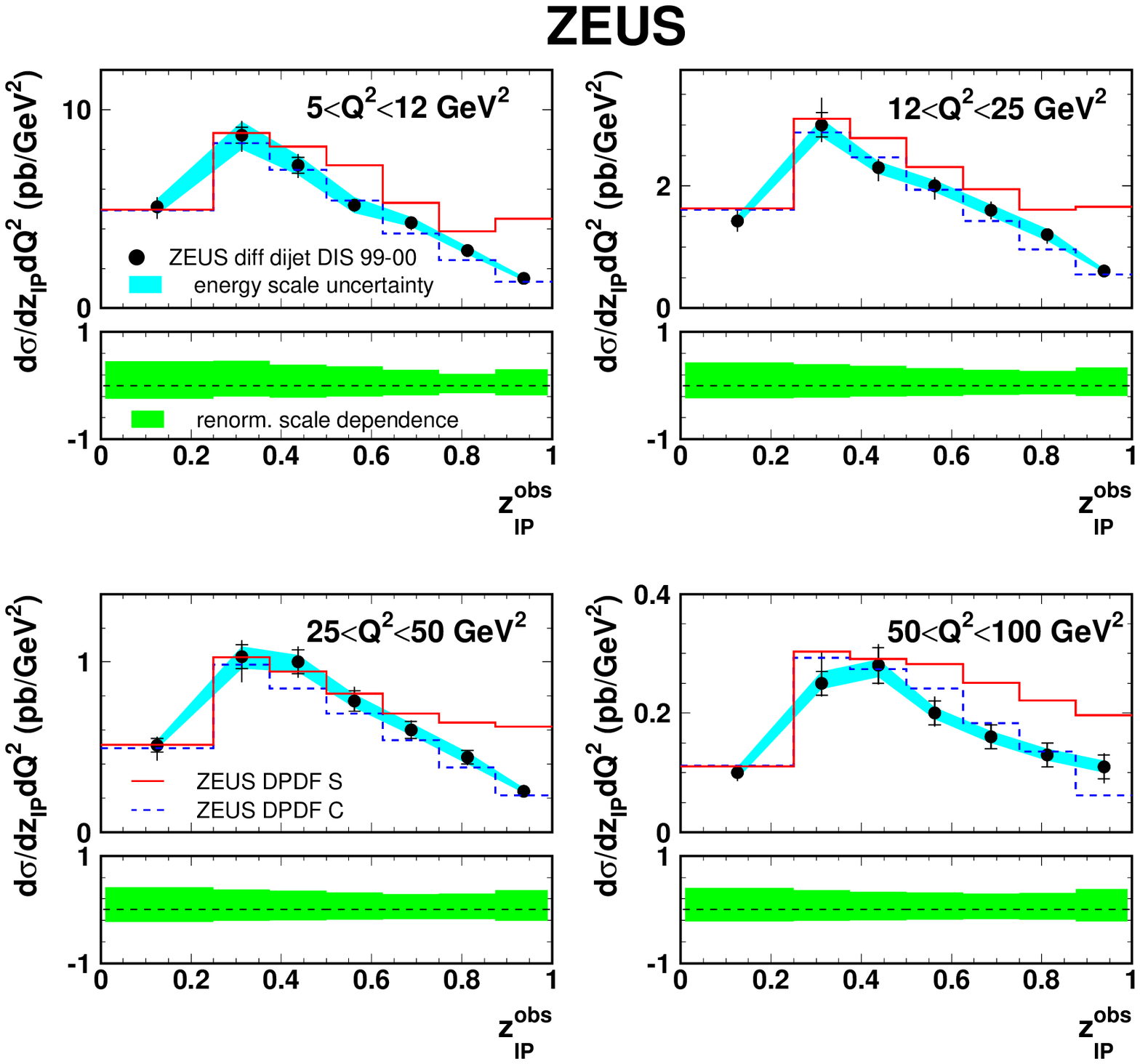}
\includegraphics[width=0.49\textwidth]{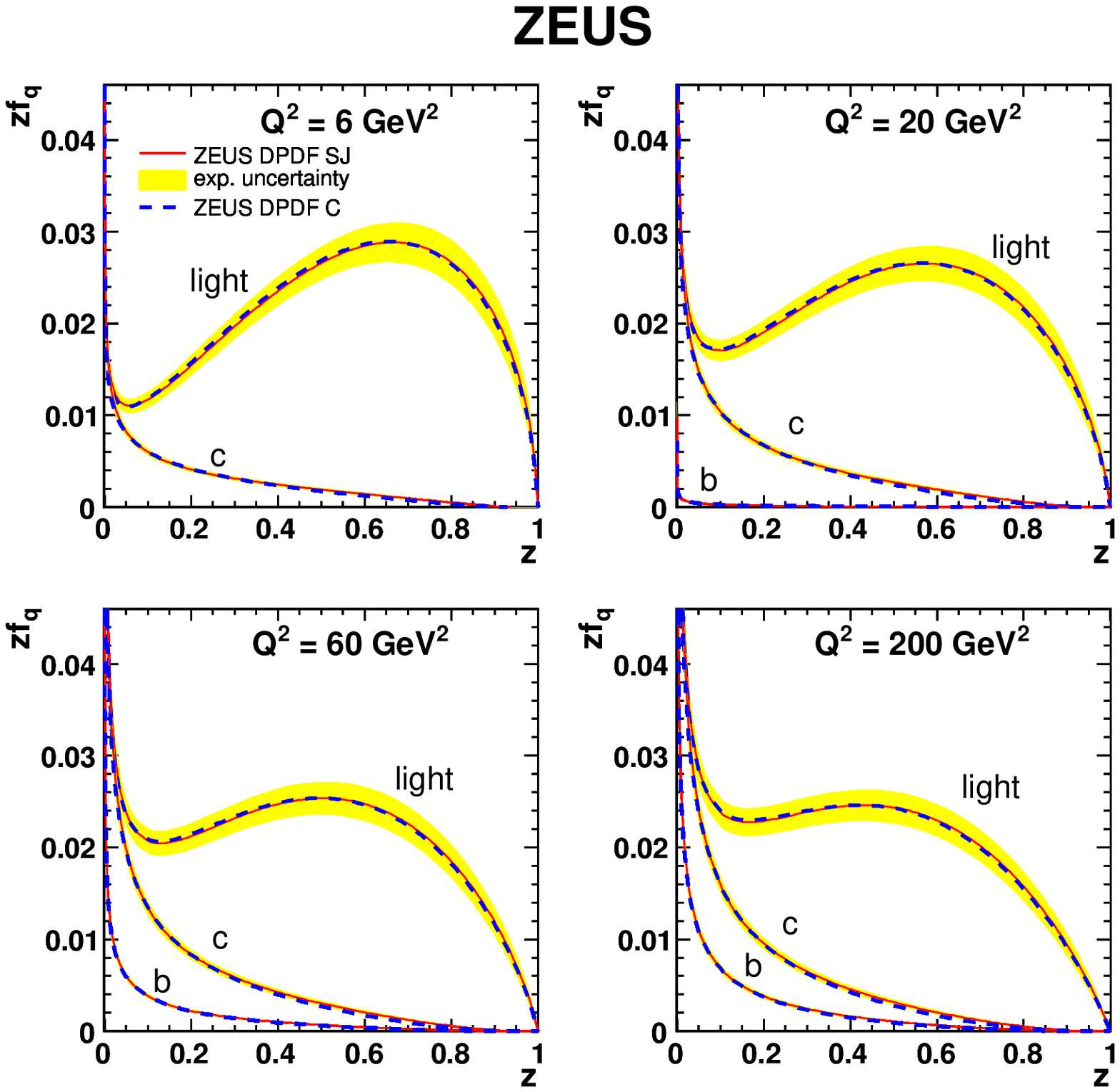}
\caption{(Left) Comparison of DPDF fits to jet data, when not used in the fit.  (Right) 
Resulting parton densities when including jet data into the fit along with inclusive 
diffractive data.}
\label{fig:dijets-dis}
\end{center}
\end{figure}

\vspace{-0.3cm}
The two fits to the inclusive data only, fits S and C gave similar distributions 
for the quark densities with small uncertainties propagated from the experimental data.  
The gluon density was very different between the two, particularly at 
high $z$ and low $Q^2$.  As the production of dijets in diffractive DIS is directly 
sensitive to the gluon content of the diffractive exchange, data on jet production 
were included.  The fit, SJ, to the dijet data was good (see 
Fig.~\ref{fig:dijets-dis}) with a similarly good description of the inclusive diffractive 
data.  The gluon density from fit SJ, with comparable uncertainties 
to that of the quark density, is similar to that of fit C and fit S is ruled out.

\section{Comparison of fits with other data}

\begin{figure}[htp]
\begin{center}
\includegraphics[width=0.49\textwidth]{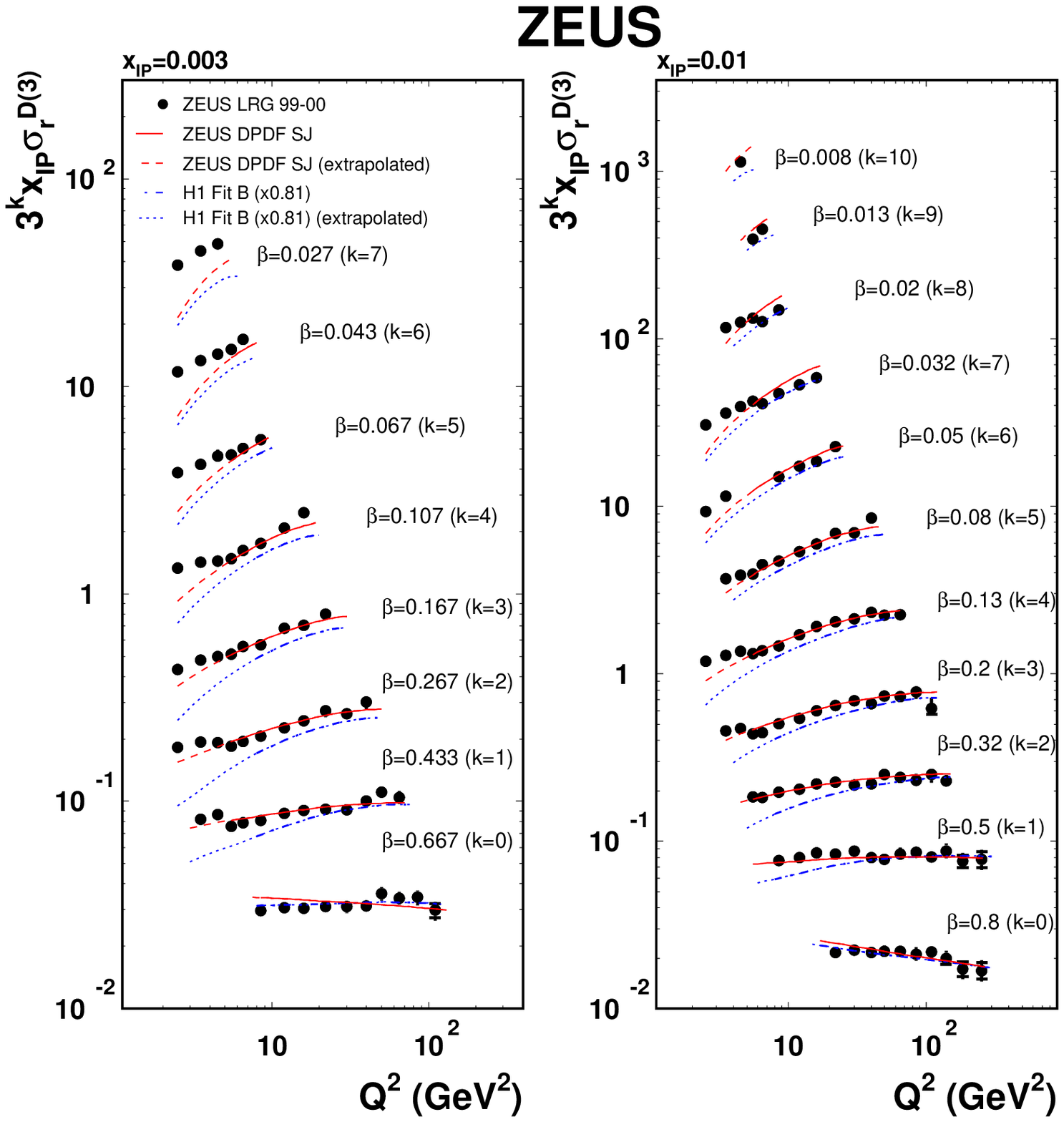}\hspace{0.5cm}
\includegraphics[width=0.44\textwidth]{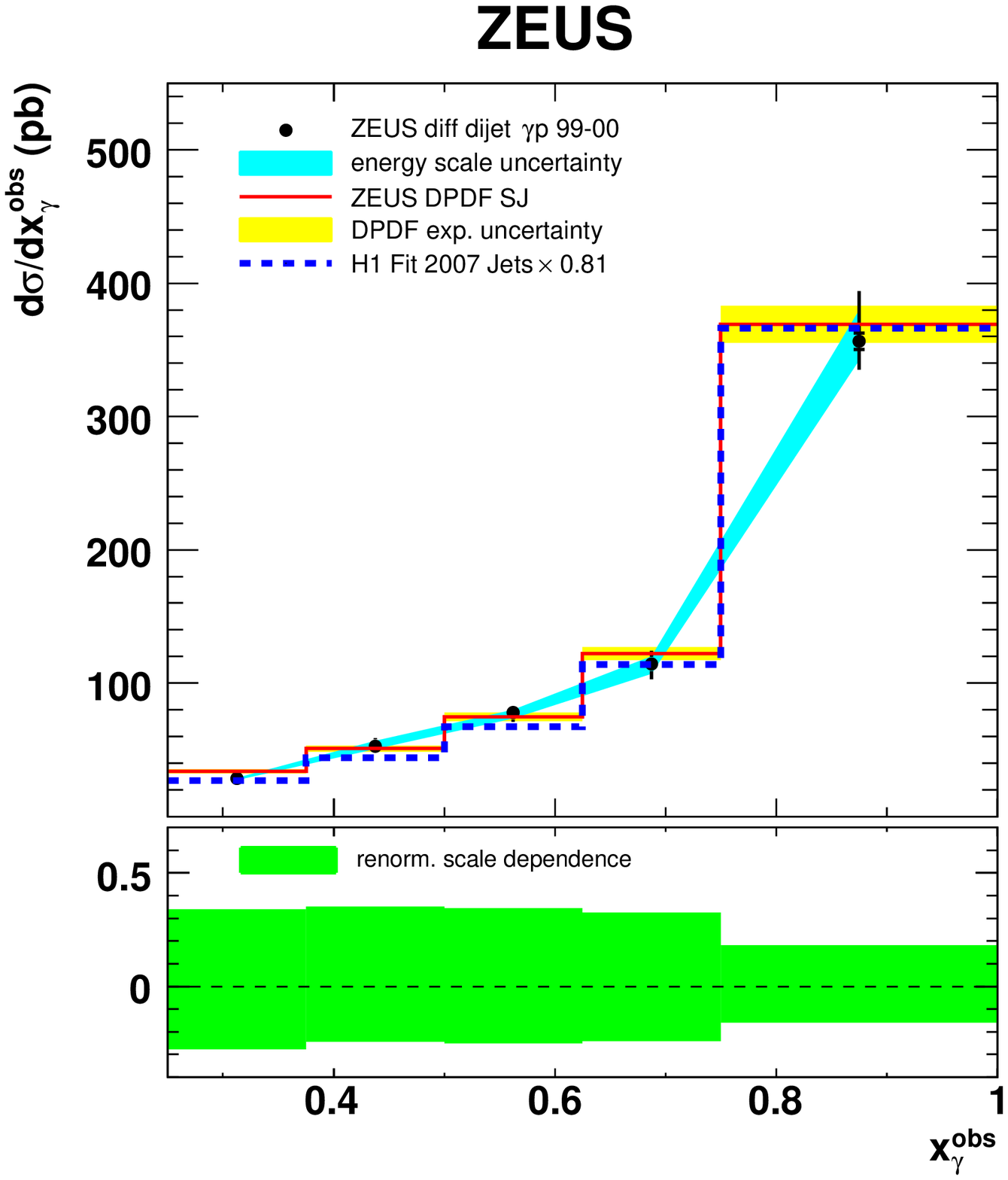}
\caption{(Left) Comparison of ZEUS DPDF fit to the ZEUS data along with the DPDF from the H1, 
Fit B.  (Right) Comparison of ZEUS DPDF fit with dijet photoproduction data.  The fractional 
uncertainty due to the renormalisation scale uncertainty is also shown.}
\label{fig:h1}
\end{center}
\end{figure}

Figure~\ref{fig:h1} shows the fit SJ compared to the inclusive diffractive data 
used in the fit along with the H1 DPDF, Fit B~\cite{H1DPDF1} which was fit to a 
similar measurement from the H1 Collaboration.  The H1 fit uses the fixed-flavour 
number scheme for heavy quarks, data with \mbox{$Q^2 > 8.5$\,GeV$^2$}, and were scaled to 
account for the different definition of the mass of the nucleon; the two fits are 
otherwise similar in spirit.  For 
$\beta < 0.2$, the shapes of the two fits are similar and differ 
by 10\%, consistent with the relative normalisation uncertainty.  In regions where the 
fits are extrapolated, they often deviate and also at higher $\beta$ which reflects the 
consistency of the two data sets.

Comparison of DPDFs extracted in DIS with jet photoproduction data, in which the scattered 
electron escapes detection down the beam-pipe provides a powerful test of the validity 
of factorisation.  In jet photoproduction, two processes can be considered :  the direct 
process where all the photon's energy participates in the hard interaction and is similar 
to DIS; and resolved processes in which the photon develops a partonic structure and the 
$\gamma P$ collision is similar to a hadron-hadron collision.  A measure of the fraction 
of photon's energy participating in the hard direction, $x_\gamma^{\rm obs}$, is shown in 
Fig.~\ref{fig:h1}.  The DPDF convoluted with the matrix-element calculation gives a good 
description of this variable with no discernible difference in the description for 
high-$x_\gamma^{\rm obs}$ (direct) and low-$x_\gamma^{\rm obs}$ (resolved) processes.  The 
DPDFs are also able to describe charm production in DIS~\cite{fit-paper}.

\section{Summary}

A QCD fit has been performed to inclusive and dijet diffractive deep inelastic scattering 
data.  The data are well described by this fit and the quark densities (from the inclusive 
data) and gluon densities (from the jet data) are well constrained.  The fit has to be 
performed for data with $Q^2 > 5$\,GeV$^2$ with data and fit deviating below this value; 
this may indicate a fundamental limit in perturbative QCD.  The DPDFs can successfully 
predict independent measurements : charm production in deep inelastic scattering and jet 
photoproduction.

The resultant DPDFs are similar to those extracted previously from H1, differing by about 
10\%.  Clearly these improved DPDFs will not improve the description of Tevatron data 
in which the prediction is an order of magnitude too high.  However, jet photoproduction 
data where a part of the cross section can be considered as a hadron-hadron interaction, 
and hence lead to secondary interactions, is well described by the extracted DPDFs.  This 
does not imply a contradiction given the differences in a hadronic-photon and a hadron 
collision, such as the ``point-like'' component of the photon structure, and the different 
energy scales involved.  Further improvements will be achieved through higher-precision 
measurements 
and understanding differences between the two collaborations and ultimately combining the 
data.  This will put predictions of Higgs production at the LHC on a surer footing, but 
the reliability of these predictions is a moot point.

\end{document}